# Vertical Clustering of 3D Elliptical Helical Data

Wasantha Samarathunga[1], Masatoshi Seki[2], Hidenobu Saito[3], Ken Ichiryu[4], Yasuhiro Ohyama[5]

[1,5]Tokyo University of Technology
Tokyo, Japan

[2,3,4]Mechatronics Research Laboratory, Kikuchi Manufacturing Co. Ltd.
Tokyo, Japan

*Abstract*— **This research proposes an effective vertical clustering strategy of 3D data in an elliptical helical shape based on 2D geometry. The clustering object is an elliptical cross-sectioned metal pipe which is been bended in to an elliptical helical shape which is used in wearable muscle support designing for welfare industry. The aim of this proposed method is to maximize the vertical clustering (vertical partitioning) ability of surface data in order to run the product evaluation process addressed in research [2]. The experiment results prove that the proposed method outperforms the existing threshold no of clusters that preserves the vertical shape than applying the conventional 3D data. This research also proposes a new product testing strategy that provides the flexibility in computer aided testing by not restricting the sequence depending measurements which apply weight on measuring process. The clustering algorithms used for the experiments in this research are self-organizing map (SOM) and K-medoids.**

I. INTRODUCTION

Computer Aided Testings of geometrically complicated shapes is very challenging task in many manufacturing fields. The resent development of 3D scanning and 3D printing brings definite advantages to industry. On the other hand rapid development on scanning precision brings too much excessive data for analysing process which slows down computation process. The effective combination of existing computer science knowledge is expected to reduce this overhead and contribute toward integrity to a certain degree. Our contribution is in machatronics for welfare industry.

In this paper we propose a novel vertical data clustering strategy. This method will be used as a pre-process to welfare product evaluation model originally described in figure 5 of research [2], which is cited in this paper as cross section-wise product evaluation procedure in figure 2. We propose that this 3D vertical clustering method could be used in other research areas as well. Research [1] briefly describes the overall machining project.

Self-organizing map (SOM) is a popular type of artificial neural networks (ANN), first introduced by Finnish professor Teuvo Kohonen. SOM is using unsupervised learning to produce low dimensional outputs, in other words maps. This method is very useful as a clustering algorithm to address nonlinear data as this algorithm clusters equally distanced data into same cluster. A successful example of SOM is described in [4].

K-medoids is an efficient clustering algorithm originally proposed by Leonard Kaufman and Peter J. Rousseeuw in 1987 as described in [3]. k-medoids is known to be robust to noise and outliers over famous k-means as it minimizes a sum of pair wise dissimilarities instead of a sum of squared Euclidean distances. The term medoid here can be explained as the object of a cluster, whose average dissimilarity to all the objects in the cluster is minimal, i.e. cluster centers are centrally located either one of clustering data.

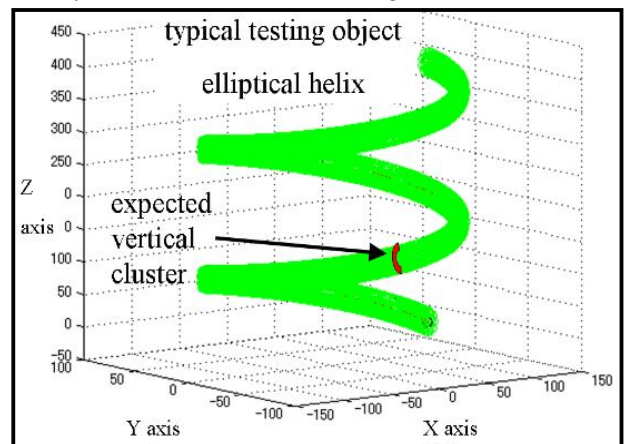
Figure 1. Vertical clustering objective

In order to proceed with the cross section-wise product evaluation for surface data as illustrated in figure 2, the resulting vertical clusters are expected to be as slim as possible.

In the following sections of this paper we address this vertical clustering in details and the relevant threshold values. The threshold number denotes here as the number of maximum clusters that preserves the resulting shape to be vertical, and with exceeding which the shape is not vertical anymore. However this number is not in absolute precision, rather rough evaluation measure due to the fact that results of clustering slightly improve with recursive trainings.

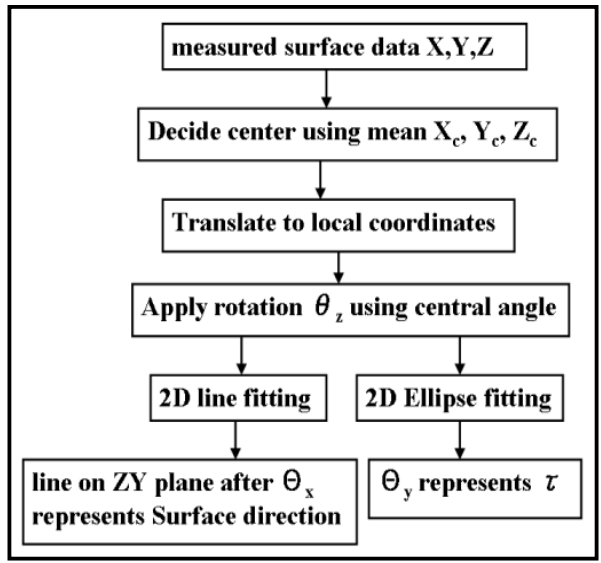

Figure 2. Cross section-wise product evaluation procedure

## II. CLUSTERING WITH 3D DATA

The following figures illustrate clustering results using 3D data for SOM and K-medoids algorithms. Figure 3 illustrates the K-medoids clustering preserves the vertical feature roughly in a 15 group clustering. Figure 4 illustrates the 20 group clustering for K-medoids. This means the threshold is between 15 to 20 groups. In figures 3, 4 and 5 "group no:number of members" is added to explain quality of clustering. SOM usually produce similarity than K-medoids.

SOM is able to manage and return 23 groups that vertical features are preserved in an expected number is originally set to be of 25, where 2 groups were empty. This allows us to assume the threshold to be roughly 23, which slightly better than K-medoids. But trainings of SOM are computationally expensive task. Figure 5 illustrates SOM 23 groups.

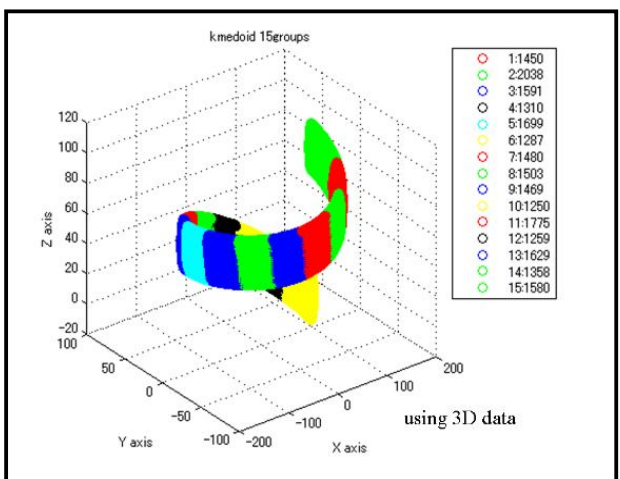

Figure 3. K-medoids 15 group clustering

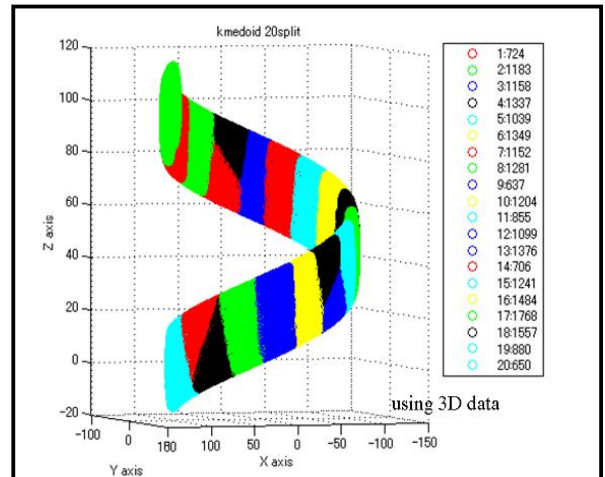

Figure 4. K-medoids 20 group clustering

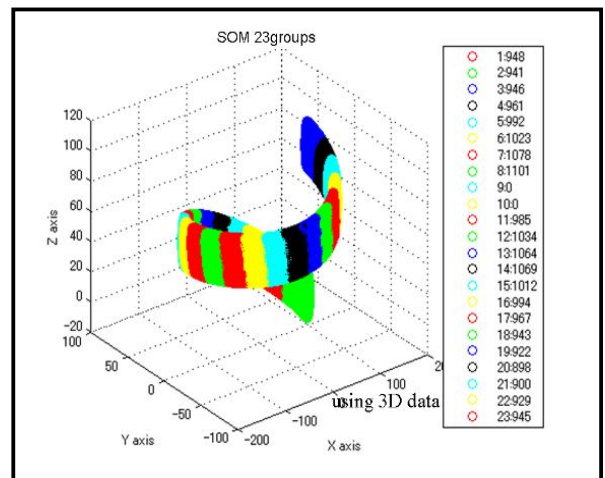

Figure 5. SOM 23 group clustering

Above results explains the vertical clustering ability with the original 3D dataset is very limited. The results are not accurate enough to proceed to the cross section-wise product evaluation procedure in figure 2.

It should be mentioned that the above experiments were performed using single turned elliptical helix to avoid confusion. Therefore it is stated that a turn-wise split is a pre-requisite even in the following alternative method. The following figure 6 illustrates this turn-wise split.

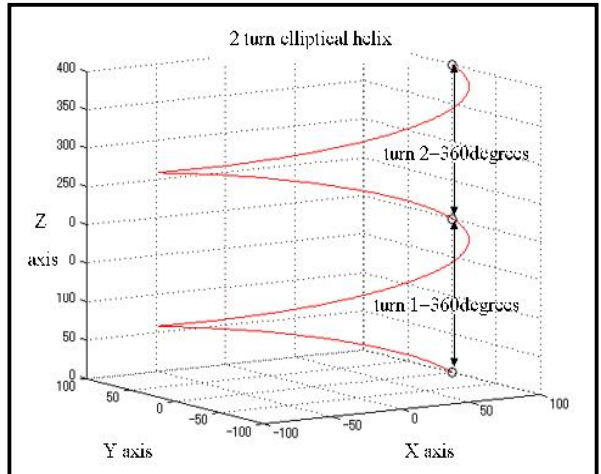

Figure 6. Turn-wise split of helix before clustering

III. PROPOSED METHOD

To achieve higher threshold value and prepare data for cross section-wise product evaluation procedure, we propose the following new strategy as illustrated in following figure 7.

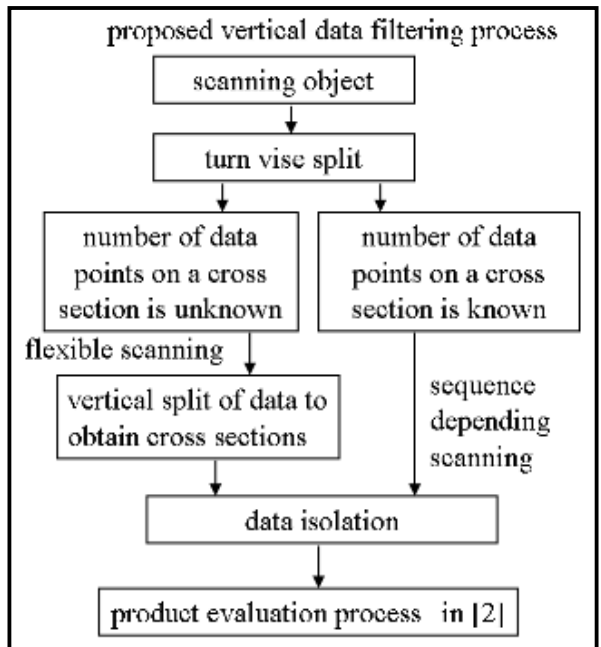

Figure 7. Proposed method

This method brings two options to proceed with. The former denotes sequence depending scanning and latter denotes flexible scanning. Both will be explained in following sections.

Due to technical merit we adopt the flexible scanning method which requires 3D vertical clustering which is more precise than in previous section follows by data isolation.

IV. SEQUENCE DEPENDING SCANNING

Sequence depending scanning assumed that number of measurement around a certain cross-section of the surface to be a constant as illustrated in figure 8. The advantage of this method is easiness of labeling each cross-section as a single group. This avoids the problem of vertical clustering and relevant cluster accuracy issue.

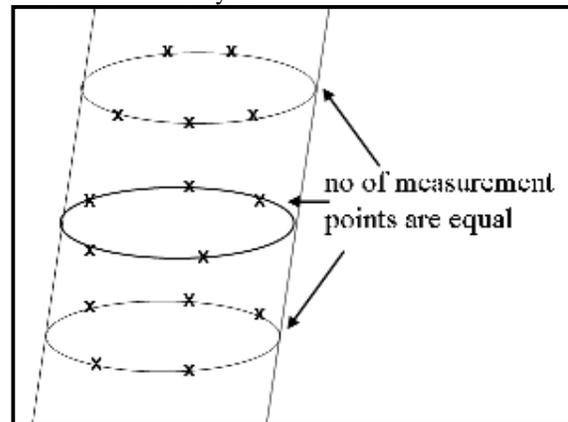

Figure 8. Sequence depending scanning

How ever this applies measuring restrictions, i.e. user has to make sure that the number of measurement is always constant. Further this method is quite impracticable for asynchronous scanning methods. Many ongoing 3D scannings developments are asynchronous, as the scanning are done on one at a time basis followed by a later on merging several scanned images to combine into a 3D model.

Since this is quite straightforward method for rest of computation process as with a known data group programmer only has to mask the data and isolate it. Hence during this paper this method is considered as less important method, but not denies the validity.

V. FLEXIBLE SCANNING

For a more generalized platform and as a good research problem it can be considered that scanning is done asynchronously and the number of data on a cross-section is not certain. In this case some sort of a clustering is the option to proceed with and clustering accuracy will become an issue.

In our case we need our data groups to be in isolated slim surfaces as illustrated in figure 1, i.e. vertical clustering. We propose use of 2D data (X and Y data only) for clustering, followed by merge the resulting clusters to original 3D data and proceed to data isolation process. This proposed method is illustrated in the following figure 9.

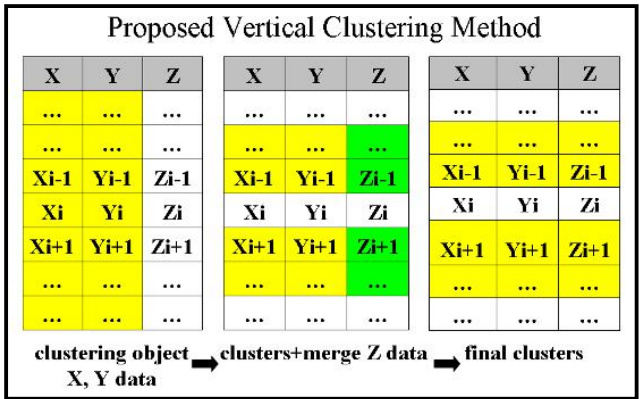

Figure 9. Flexible Scanning

The following are experiment results with new method. Again we use both algorithms SOM and K-medoids. The following figure 10 illustrates 72 groups of vertical clusters for K-medoids and figure 11 illustrates 72 groups of vertical clusters for SOM.

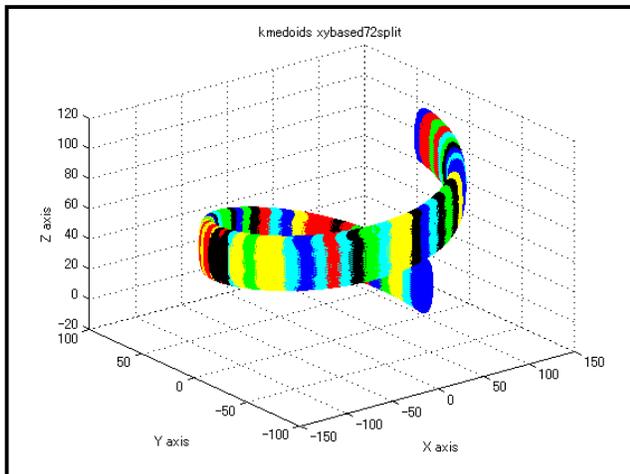

Figure 10. K-medoids 72 vertical clusters

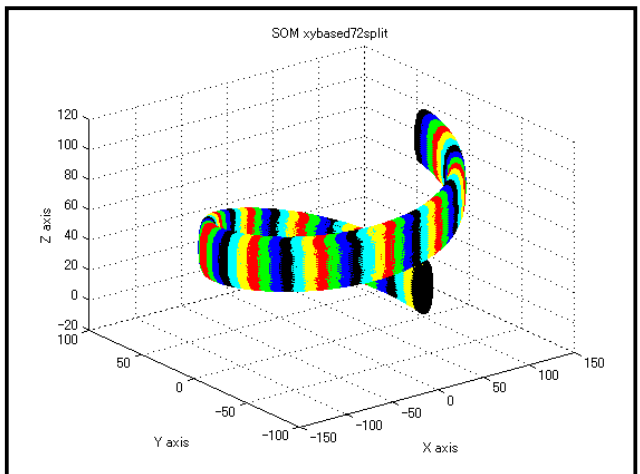

Figure 11. SOM 72 vertical clusters

With the above results it can be stated that vertical clustering ability is improved as threshold values shown in following table 1.

TABLE 1.
THRESHOLD NUMBERS

|  | K-medoids | SOM |
|---|---|---|
| conventional 3D | 15～20 | 23 |
| proposed method | 72～ | 72～ |

Although the upper limit of above threshold numbers is still unknown due to slow development process, it can be stated that there is a definite improvement using the new strategy.

## VI. CONCLUSIONS

The preliminary 3D oriented clustering experiments shows that SOM works better than K-medoids in preserving the vertical shape in clustering with higher threshold number, but met with limitations to proceed with due to excessive information related to Z axis.

With the proposed strategy of using 2D data for clustering and then merge the clusters to original 3D data proved to increase the vertical clustering ability as current thresholds recorded to be of 72 (means 5 degree interval).

It could be stated that further improvement is theoretically possible, but development with SOM is too slow. Also the actual product is also still due.

ACKNOWLEDGMENT

The authors express their sincere gratitude to Kikuchi Seisakusho Co. Ltd. at Hachioji-City, Tokyo for the financial and technical support for this project.